\documentclass[preprint]{ptephy_v1}
\usepackage{amssymb}
\usepackage{graphicx}
\numberwithin{equation}{section}

\newcommand{\cb}{\bar{c}}
\newcommand{\lb}[1]{\label{#1}}
\newcommand{\cl}{\mathcal{L}}
\newcommand{\bb}{\bar{B}}

\newcommand{\vp}{\varphi}
\newcommand{\ptl}{\partial}

\def\vec#1{\mbox{\boldmath $#1$}}
\newcommand{\ct}{\tilde{C}}
\newcommand{\hh}{\mathcal{H}}
\newcommand{\cvp}{\tilde{\varphi}}
\newcommand{\cc}{\mathcal{C}}

\begin{document}

\title{Massive dual gauge field and confinement in Minkowski space : Magnetic charge
}


\author{\name{\fname{Hirohumi} \surname{Sawayanagi}}{1}}

\address{\affil{1}{National Institute of Technology, Kushiro College, Kushiro, 084-0916, Japan}
\email{sawa@kushiro-ct.ac.jp}}


\begin{abstract}
     Gauge field configuration for a magnetic monopole and its dual configuration 
are studied in SU(2) gauge theory.  We present a relation between the monopole field 
and its dual field.  Since these fields can become massive, their massive Lagrangians 
are derived.  In the dual case, an additional term appears.  We show this term is 
necessary to produce a linear potential between a monopole charge and an antimonopole charge.  
\end{abstract}

\subjectindex{B0,B3,B6}

\maketitle

\section{Introduction}

     Magnetic monopoles are considered to play an important role in quark confinement.  
Models based on the dual superconductor require monopoles and their condensation (see, e.g., \cite{rip}).  
In the Zwanziger's formulation \cite{zwa}, 
two gauge fields, namely the usual gauge field and the dual gauge field, are used.  
To describe monopole condensation, a monopole field is introduced.  Its vacuum expectation value (VEV) 
makes the dual gauge field massive.  This massive field leads to 
a linear potential between a quark and an anti-quark.  

     In the extended QCD \cite{cho1}, a unit color vector $\hat{n}^A(x)$ 
in the internal space \cite{cho,kon} appears.  Non-Abelian magnetic potential 
defined by 
$C_{\mu}^A =  -(\hat{n}\times \ptl_{\mu}\hat{n})^A/g$
describes the non-Abelian monopole \cite{wy}.  In Ref.~\cite{hs}, we studied the SU(2) gauge theory 
in a nonlinear gauge, and derived the extended QCD with the massive magnetic potential $C_{\mu}^A$.  
In this paper, we study the relation between the Abelian magnetic potential and its dual potential.  
Based on this relation, we consider the confinement of magnetic charges.  

     In Sect.2, we introduce the dual magnetic potential for magnetic monopoles.  
Next, by using the dual magnetic potential, we rewrite the Abelian part of the SU(2) Lagrangian 
with the monopole field.  Massless magnetic potential is considered in Sect. 3, and massive magnetic potential 
is studied in Sect. 4.  
In Sect. 5, SU(2) gauge theory in the low energy region is studied.  
Because of the ghost condensation \cite{sch, ks} and the condensate $\langle A_{\mu}^+A^{-\mu}\rangle$, 
the massive magnetic potential appears in this region \cite{hs}.  
Applying the result in Sect. 4, we obtain the low energy SU(2) Lagrangian with the dual 
magnetic potential.  In Sect. 6, using the Lagrangian in Sect. 5, it is shown that the 
static potential between a magnetic monopole and an anti-monopole is a 
linear confining potential.  Section 7 is devoted to summary.  
In Appendix A, notations are summarized.  For both the massless case and the massive case, 
the monopole solutions and their dual solutions for a static magnetic charge are given in Appendix B.  
We also show that the relation between the Abelian magnetic potential and its dual potential, 
which is presented in the Sect. 2, is satisfied for the Dirac monopole.  
To obtain the Lagrangian in Sect. 5, the ghost condensation is necessary.  
In Ref.~\cite{hs2}, we have shown that it happens in Euclidean space.  
In Appendix C, we show it in Minkowski space.

\section{Dual magnetic potential}

     Let us consider a space-like gauge field $\ct_{\mu}$ which satisfies the equation of motion 
\begin{equation}
   \ptl^{\nu}H_{\mu\nu}=\epsilon_{\mu\nu\alpha\beta}\ptl^{\nu}\frac{n^{\alpha}}{n^{\rho} \ptl_{\rho}}k^{\beta}, \quad 
   H_{\mu\nu}=(\ptl\wedge \ct)_{\mu\nu},   \lb{201}
\end{equation}
where $(\ptl\wedge \ct)_{\mu\nu}=\ptl_{\mu} \ct_{\nu}-\ptl_{\nu}\ct_{\mu}$, the space-like vector $n^{\alpha}$ satisfies 
$n_{\alpha}n^{\alpha}=-1$, and the magnetic current $k^{\beta}$ satisfies $\ptl_{\beta}k^{\beta}=0$.  
We call $\ct_{\mu}$ magnetic potential.  The dual field strength is defined by 
\begin{equation}
     \hh^{\mu\nu}=\frac{1}{2}\epsilon^{\mu\nu\alpha\beta}H_{\alpha\beta}=\epsilon^{\mu\nu\alpha\beta}\ptl_{\alpha}\ct_{\beta}.  \lb{202}
\end{equation}
Now we introduce a dual magnetic potential $\cc_{\mu}$.  
Using the formula 
$\epsilon^{\alpha\beta\rho\sigma}\epsilon_{\rho\sigma\mu\nu}=-2(\delta^{\alpha}_{\mu}\delta^{\beta}_{\nu}
     -\delta^{\alpha}_{\nu}\delta^{\beta}_{\mu})$,
\footnote{We employ the metric $g_{\mu\nu}=\mathrm{diag}(1,-1,-1,-1)$, and the anti-symmetric $\epsilon$ symbol 
with $\epsilon^{0123}=1$.  The formulae related to $\epsilon^{\mu\nu\alpha\beta}$ are summarized in Appendix A.} 
we obtain 
\begin{equation}
     H_{\mu\nu}H^{\mu\nu}=-\hh_{\rho\sigma}\hh^{\rho\sigma}.  \lb{203}
\end{equation}
Therefore, if we define $\cc_{\mu}$ as $\hh^{\mu\nu}=(\ptl\wedge \cc)^{\mu\nu}$, 
the kinetic term for $\cc_{\mu}$ has the wrong sign \cite{cho}.  We change this relation to 
\begin{equation}
     \hh^{\mu\nu}=(\ptl\wedge \cc)^{\mu\nu}+\Lambda^{\mu\nu}. \lb{204}
\end{equation}
If we impose the conditions 
\begin{equation}
     \ptl_{\mu}\hh^{\mu\nu}=\epsilon^{\mu\nu\alpha\beta}\ptl_{\mu}\ptl_{\alpha}\ct_{\beta}=0, \quad
     \ptl_{\nu}\hh^{\mu\nu}=\epsilon^{\mu\nu\alpha\beta}\ptl_{\nu}\ptl_{\alpha}\ct_{\beta}=0,  \lb{205}
\end{equation}
$\Lambda^{\mu\nu}$ must satisfy 
\begin{equation}
     \ptl_{\mu}(\ptl\wedge \cc)^{\mu\nu}+\ptl_{\mu}\Lambda^{\mu\nu}=0,\quad 
     \ptl_{\nu}(\ptl\wedge \cc)^{\mu\nu}+\ptl_{\nu}\Lambda^{\mu\nu}=0.  \lb{206}
\end{equation}
Eq.(\ref{206}) holds if we choose 
\begin{equation}
     \Lambda^{\mu\nu}= -\frac{n^{\mu}}{n^{\rho}\ptl_{\rho}}\ptl_{\sigma}(\ptl\wedge \cc)^{\sigma\nu}
     +\frac{n^{\nu}}{n^{\rho}\ptl_{\rho}}\ptl_{\sigma}(\ptl\wedge \cc)^{\sigma\mu}.  \lb{207}
\end{equation}
Thus we obtain the relation 
\begin{equation}
     \hh^{\mu\nu}=\epsilon^{\mu\nu\alpha\beta}\ptl_{\alpha}\ct_{\beta}=
     (\ptl\wedge \cc)^{\mu\nu}-\frac{n^{\mu}}{n^{\rho}\ptl_{\rho}}\ptl_{\sigma}(\ptl\wedge \cc)^{\sigma\nu}
     +\frac{n^{\nu}}{n^{\rho}\ptl_{\rho}}(\ptl_{\sigma}(\ptl\wedge \cc)^{\sigma\mu}.  \lb{208}
\end{equation}
From Eq.(\ref{208}), we can write $H^{\mu\nu}$ as 
\begin{align}
 H^{\mu\nu}&= (\ptl\wedge \ct)^{\mu\nu}=h_{1}^{\mu\nu}+h_{2}^{\mu\nu}, \nonumber \\
 h_{1}^{\mu\nu}&=-\epsilon^{\mu\nu\alpha\beta}\ptl_{\alpha}\cc_{\beta}, \quad 
 h_{2}^{\mu\nu}=\epsilon^{\mu\nu\alpha\beta}\frac{n_{\alpha}}{n^{\rho}\ptl_{\rho}}\ptl^{\sigma}(\ptl\wedge \cc)_{\sigma\beta}
 =-\frac{1}{2}\epsilon^{\mu\nu\alpha\beta}\Lambda_{\alpha\beta}.  
\lb{209}
\end{align}
This expression will be used in the following sections.  

     We note $H_{\mu\nu}$ is invariant under the transformations 
\[ \ct_{\mu} \to \ct_{\mu}+ \ptl_{\mu} \varepsilon, \quad \cc_{\mu} \to \cc_{\mu}+ \ptl_{\mu} \vartheta. \]
If we choose the gauges
\begin{equation}
     n_{\mu}\ct^{\mu}=0, \quad  n_{\mu}\cc^{\mu}=0,  \lb{210}
\end{equation}
Eq.(\ref{209}) is solved as 
\begin{equation}
     \ct^{\nu}=\epsilon^{\nu\mu\alpha\beta}\frac{n_{\mu}\ptl_{\alpha}}{n^{\rho}\ptl_{\rho}}\cc_{\beta}.  \lb{211}
\end{equation}

     In Appendix B, as an example, $\ct_{\mu}$ and $\cc^{\mu}$ 
for a static magnetic charge $ k^{\beta}\propto \delta^{\beta}_0 \delta(x)\delta(y)\delta(z)$
are presented.  The magnetic potential $\ct_{\mu}$ is the Dirac monopole and its dual potential $\cc^{\mu}$ 
is the Coulomb potential.  
Eqs.(\ref{205}) and (\ref{208}) are fulfilled by these potentials, and the term 
$\Lambda^{\mu\nu}$ represents the Dirac string.

\section{Abelian part of the SU(2) gauge theory}

  We consider the SU(2) gauge theory with structure constants $f^{ABC}$.  
Using the notations 
\[ F\cdot G=F^AG^A, \quad (F\times )^{AB}=f^{ACB}F^C, \quad
(F\times G)^A=f^{ABC}F^BG^C, \quad 
A=1,2,3, \]
the Lagrangian 
\begin{equation}
     \cl_{\mathrm{inv}}(A)=-\frac{1}{4}G_{\mu\nu}\cdot G^{\mu\nu},\quad 
     G_{\mu\nu}^A=(\ptl\wedge A^A)_{\mu\nu}+g(A_{\mu}\times A_{\nu})^A  \lb{301}
\end{equation}
contains the Abelian part 
\begin{equation}
     \cl_{\mathrm{Abel}}(A)=-\frac{1}{4}(\ptl\wedge A^3)_{\mu\nu}(\ptl\wedge A^3)^{\mu\nu}.    \lb{302}
\end{equation}
To incorporate the magnetic potential, we divide the gauge field $A_{\mu}^3$ into a classical part $b_{\mu}^A$ 
and a quantum part $a_{\mu}^A$ as 
\begin{equation}
     A_{\mu}^A=b_{\mu}^A + a_{\mu}^A,\quad b_{\mu}^A=\ct_{\mu}\delta^{A}_{3}.  \lb{303}
\end{equation}
For simplicity, we use the notation $(F+H)_{\mu\nu}(F+H)^{\mu\nu}=(F + H)^2$.  Then, $\cl_{\mathrm{Abel}}(A)$ becomes 
\begin{equation}
     \cl_{\mathrm{Abel}}(b+a)=-\frac{1}{4}(F + H)^2, 
     \quad F_{\mu\nu}=(\ptl\wedge a^3)_{\mu\nu},\quad H_{\mu\nu}=(\ptl\wedge \ct)_{\mu\nu}.  \lb{304}
\end{equation}

     Next we introduce the magnetic current $k^{\beta}$.  To reproduce the equation of motion (\ref{201}), 
we consider the Lagrangian 
\footnote{The field strength $H+h_3$ is the Zwanziger's field strength $F=(\ptl \wedge A)-(n\cdot \ptl)^{-1}(n\wedge j_g)^d$ 
in Ref.~\cite{zwa}.}
\begin{equation}
     \cl'_{\mathrm{Abel}}(b+a)=-\frac{1}{4}(F + H + h_3)^2, \quad 
      h_{3}^{\mu\nu}=-\epsilon^{\mu\nu\alpha\beta}\frac{n_{\alpha}}{n^{\rho}\ptl_{\rho}}k_{\beta}.  \lb{305}
\end{equation}
Then, using Eq.(\ref{209}), we rewrite Eq.(\ref{305}) as 
\begin{equation}
     \cl'_{\mathrm{Abel}}(b+a)=-\frac{1}{4}(F + h_2 + h_3)^2
      -\frac{1}{4}h_{1\mu\nu}(h_1^{\mu\nu}+2h_2^{\mu\nu}+2h_3^{\mu\nu}),   \lb{306}
\end{equation}
where $\int dx F_{\mu\nu}h_1^{\mu\nu}=0$ has been used.  
As we stated in Eq.(\ref{203}), the part 
\begin{equation}
    -\frac{1}{4}h_{1\mu\nu}h_1^{\mu\nu}=\frac{1}{4}(\ptl\wedge \cc)^{2}  \lb{307}
\end{equation}
gives the kinetic term with the wrong sign.  However, using the current conservation $\ptl_{\mu}k^{\mu}=0$, we obtain  
\begin{equation}
     -\frac{1}{4}h_{1\mu\nu}(2h_2^{\mu\nu}+2h_3^{\mu\nu})=-\frac{1}{2}(\ptl\wedge \cc)^{2} 
     -\cc_{\mu}k^{\mu}.  \lb{308}
\end{equation}
Thus we find the cross term $2h_{1\mu\nu}h_2^{\mu\nu}$ changes the sign of the kinetic term for $\cc_{\mu}$,
\footnote{Similar result is found in the study of the interaction energy of two magnetic monopoles.  
The cross term of the Coulomb part and the string part changes the sign of the energy of the Coulomb interaction \cite{cgpz}.}
and 
$-\frac{1}{4}h_{1\mu\nu}(h_1^{\mu\nu}+2h_2^{\mu\nu})$ yields the correct kinetic term.  Thus we obtain 
\begin{equation}
     \cl'_{\mathrm{Abel}}(b+a)=-\frac{1}{4}(F + h_2 + h_3)^2
      -\frac{1}{4}(\ptl\wedge \cc)^{2} 
     -\cc_{\mu}k^{\mu},   \lb{309}
\end{equation}

     Now we neglect the quantum part $F_{\mu\nu}$.  Then the classical solution $\cc_{\nu}$ must satisfy 
the equation of motion 
\begin{equation}
     -\frac{1}{2}(h_2^{\mu\rho} + h_3^{\mu\rho})\frac{\delta h_{2\mu\rho}}{\delta \cc^{\nu}}
     +\ptl^{\mu}(\ptl\wedge \cc)_{\mu\nu} - k_{\nu}=0.  \lb{310}
\end{equation}
However Eq.(\ref{310}) is satisfied by 
\begin{equation}
     \ptl^{\mu}(\ptl\wedge \cc)_{\mu\nu} - k_{\nu}=0,   \lb{311}
\end{equation}
because Eq.(\ref{311}) leads to $h_2^{\mu\nu} + h_3^{\mu\nu}=0$.  If we insert $h_2^{\mu\nu} + h_3^{\mu\nu}=0$ 
into the Lagrangian (\ref{309}), the term $h_2^{\mu\nu}$, which is related to the Dirac string $\Lambda_{\alpha\beta}$, 
disappears.

\section{Massive Abelian part of the SU(2) gauge theory}

     In the previous paper \cite{hs}, we have shown that there appears the mass terms 
\begin{equation}
     \cl_m = \frac{m^2}{2}\left(2a^3_{\mu}\ct^{\mu} + \ct_{\mu}\ct^{\mu}\right),  \lb{401}
\end{equation}
where the mass squared $m^2$ is defined in Eq.(\ref{505}), and the derivation of $\cl_m$ is explained briefly in Sect. 5.  
Using Eq.(\ref{211}) and integration by parts, the first term becomes 
\begin{equation}
     \int dx \ m^2 a^3_{\mu}\ct^{\mu} =-\frac{1}{2} \int dx F_{\mu\nu}h_4^{\mu\nu}, \quad 
     h_{4}^{\mu\nu}=\epsilon^{\mu\nu\alpha\beta}\frac{n_{\alpha}}{n^{\rho}\ptl_{\rho}}m^2 \cc_{\beta},  \lb{402}
\end{equation}
and the second term becomes 
\begin{equation}
     \int dx \frac{m^2}{2} \ct_{\mu}\ct^{\mu} = \int dx \frac{m^2}{2} \left[ \cc_{\mu}\cc^{\mu} + 
     \frac{n_{\beta}}{n^{\rho}\ptl_{\rho}}\cc_{\mu}(\Box \delta^{\mu}_{\nu}-\ptl^{\mu}\ptl_{\nu}) 
     \frac{n^{\beta}}{n^{\sigma}\ptl_{\sigma}}\cc^{\nu} \right],  \lb{403}
\end{equation}
where Eq.(\ref{a03}) and the gauge condition $n_{\nu}\cc^{\nu}=0$ have been used.  

If we apply Eq.(\ref{a02}) and $n_{\nu}\cc^{\nu}=0$, Eq.(\ref{403}) is rewritten as 
\begin{equation}
     \int dx \left( \frac{m^2}{2} \cc_{\mu}\cc^{\mu} - \frac{1}{4}h_2^{\mu\nu} h_{4\mu\nu} \right).  \lb{404}
\end{equation}

     Now, combining the kinetic term (\ref{305}) with the mass term (\ref{401}), we consider the Lagrangian 
$\cl_{\mathrm{mAbel}}(b+a) =\cl'_{\mathrm{Abel}}+\cl_m$.  
Applying Eqs.(\ref{309}), (\ref{402}) and (\ref{404}), it becomes 
\begin{align}
\cl_{\mathrm{mAbel}}(b+a) =& -\frac{1}{4}(F + h_2 + h_3 + h_4)^2  
     -\frac{1}{4}(\ptl\wedge \cc)^{2} +\frac{m^2}{2} \cc_{\mu}\cc^{\mu} - \cc_{\mu}k^{\mu}+ \Omega , \lb{405} \\
   \Omega=& \frac{1}{4}\left(h_2^{\mu\nu} + 2h_3^{\mu\nu}+h_4^{\mu\nu}\right) h_{4\mu\nu}.  \nonumber
\end{align}
This is the massive Abelian part expressed by the dual magnetic potential $\cc_{\mu}$.  

     Since the relation 
\[   \int dx h_2^{\mu\rho}\frac{\delta h_{4\mu\rho}}{\delta \cc^{\nu}}=\int dx \frac{\delta h_{2}^{\mu\rho}}{\delta \cc^{\nu}}h_{4\mu\rho} \]
holds, we find 
\[ \frac{\delta \Omega}{\delta \cc^{\nu}}=
 \frac{1}{2}(h_2^{\mu\rho} + h_3^{\mu\rho}+h_4^{\mu\rho})\frac{\delta h_{4\mu\rho}}{\delta \cc^{\nu}}.  \]
Then the Lagrangian (\ref{405}) leads to the following equation of motion for $\cc_{\nu}$:  
\begin{equation}
  -\frac{1}{2}(h_2^{\mu\rho} + h_3^{\mu\rho}+h_4^{\mu\rho})\frac{\delta h_{2\mu\rho}}{\delta \cc^{\nu}}
  +(\Box +m^2) \cc_{\nu}-\ptl_{\nu}\ptl_{\mu}\cc^{\mu} - k_{\nu}=0.  \lb{406} 
\end{equation}
Eq.(\ref{406}) is satisfied by the equation of motion 
\begin{equation}
     \left[(\Box +m^2)\delta^{\mu}_{\nu} -\ptl^{\mu}\ptl_{\nu}\right]\cc^{\nu}-k^{\mu}=0,  \lb{407}
\end{equation}
because Eq.(\ref{407}) leads to $h_2^{\mu\nu}+h_3^{\mu\nu}+h_4^{\mu\nu}=0$.  For a static magnetic charge, 
a solution of Eq.(\ref{407}) is presented in Appendix B.  

     We note, using Eq.(\ref{a02}), $\Omega$ is rewritten as 
\begin{equation}
     \Omega=\cc^{\mu}\frac{m^2}{2}\frac{n_{\alpha}n^{\alpha}}{(n^{\rho}\ptl_{\rho})^2}
     \left(g_{\mu\nu}-\frac{n_{\mu}n_{\nu}}{n_{\sigma}n^{\sigma}}\right)\left[\ptl_{\lambda}(\ptl \wedge \cc)^{\lambda\nu}
     +m^2\cc^{\nu}-2k^{\nu}\right].   \lb{408}
\end{equation}

\section{SU(2) gauge theory in the low energy region}

\subsection{Derivation of the massive magnetic potential}

     In this subsection, we review the derivation of the Lagrangian with the massive magnetic potential \cite{hs}.  
     Let us consider the Lagrangian 
\begin{equation}
     \cl(b+a)=\cl_{\mathrm{inv}}(b+a) + \cl_{\vp}(a,b),  \lb{501}
\end{equation}
where $\cl_{\mathrm{inv}}(A)$ is defined in Eq.(\ref{301}), and $A_{\mu}^A=b_{\mu}^A+a_{\mu}^A$.  
In the background covariant gauge, a gauge-fixing part is chosen as 
\begin{align}
 \cl_{\vp}(a,b)=& \frac{\alpha_1}{2}B\cdot B+B\cdot [D_{\mu}(b)a^{\mu}+\vp-w ] \nonumber \\
 & + i\cb\cdot [D_{\mu}(b)D^{\mu}(b+a)+g\vp \times] c - \frac{\vp\cdot \vp}{2\alpha_2},   \lb{502}
\end{align}
where $D_{\mu}(A)=\ptl_{\mu}+gA_{\mu}\times$, $\vp^A$ is a field in the adjoint representation, and 
$w^A$ is a constant.  
If $\vp^A$ is integrated out, Eq.(\ref{502}) gives 
\begin{equation}
 \cl_{\mathrm{NL}} =  B\cdot D_{\mu}(b)a^{\mu} + i\cb\cdot[D_{\mu}(b)D^{\mu}(b+a)c]
 + \frac{\alpha_1}{2}B\cdot B + \frac{\alpha_2}{2}\bb\cdot \bb -B\cdot w,   \lb{503}
\end{equation}
where $\alpha_1$ and $\alpha_2$ are gauge parameters, and $\bb = -B+ ig\cb \times c$.  
Namely Eq.(\ref{501}) gives the Lagrangian in the nonlinear gauge \cite{bt} with the constant $w^A$.  
Although $\vp^A$ is the auxiliary field which represents $\alpha_2\bb^A$, because of 
the quartic ghost interaction $\frac{\alpha_2}{2}\bb\cdot\bb$, it acquires the VEV 
$\langle \vp^A\rangle =\vp_0\delta^{A}_{3}$ under the energy scale $\mu_0=\Lambda e^{-4\pi^2/(\alpha_2g^2)}$ \cite{hs3}
,where $\Lambda$ is a momentum cut-off.
\footnote{In Ref.~\cite{hs2}, the ghost condensation $\vp_0\neq 0$ is shown in Euclidean space.  
In Appendix C, we explain that this phenomenon happens in Minkowski space as well. }  
Substituting $\vp^A(x)=\vp_0\delta^{A}_{3}+\cvp^A(x)$ into Eq.(\ref{502}), we obtain 
\begin{align}
 \cl_{\vp}(a,b)=& \frac{\alpha_1}{2}B\cdot B 
 +B\cdot [D_{\mu}(b)a^{\mu}+\cvp]  \nonumber \\
 &+ i\cb\cdot [D_{\mu}(b)D^{\mu}(b+a) +v\times  +g\cvp \times ]c - \frac{(v+g\cvp)\cdot(v+g\cvp)}{2\alpha_2 g^2},  \lb{504}
\end{align}
where $v=g\vp_0$, and to preserve the BRS symmetry, $w^A$ is chosen as $w^{A}=\vp_0\delta^{A}_{3}$ \cite{kug,hs,hs1}.  
Since $\vp_0\delta^{A}_{3}$ selects the unbroken U(1) direction, we incorporate the Abelian monopole  
as $b_{\mu}^A=\ct_{\mu}\delta^{A}_{3}$ \cite{afg}.  
     
     When $v\neq 0$, it is known that ghost loops bring about tachyonic gluon masses \cite{hs2, dv}. 
In Ref.~\cite{hs}, we have shown that the ghost determinant $\det[D_{\mu}(b)D^{\mu}(b+a)+v\times ]$ yields the tachyonic mass terms 
\begin{equation}
 -\frac{m^2}{2} [a_{\mu}^+a^{-\mu} + a_{\mu}^3a^{3\mu}], \quad  m^2=\frac{g^2v}{32\pi},  \lb{505} 
\end{equation}
where $a_{\mu}^+a^{-\mu}=a_{\mu}^aa^{a\mu}/2\ (a=1,2)$.  
We note, contrary to the quantum part $a_{\mu}^A$, $b_{\mu}^A$ does not have tachyonic mass.  

     To avoid the tachyonic masses, we introduced the source term $M^2a^+_{\mu}a^{-\mu}$ into the Lagrangian, and 
constructed the effective potential for $\Phi=\langle a^+_{\mu}a^{-\mu}\rangle$ \cite{hs}.  However, at the lowest order, 
we can obtain the effective potential by the following simple procedure.  First add the tachyonic mass terms 
(\ref{505}) and the source term $M^2a^+_{\mu}a^{-\mu}$ to the Lagrangian.  
Next replace $a_{\mu}^+ a^{-\mu}$ to $\Phi + a_{\mu}^+ a^{-\mu}$.  Thus 
the terms which contain $\Phi$, $m^2$, or $M^2$ are 
\begin{align}
 &-V(\Phi) - \left(g^2\Phi +\frac{m^2}{2} - M^2 \right)a_{\mu}^+a^{-\mu} 
 - \left(g^2\Phi +\frac{m^2}{2} \right)a_{\mu}^3a^{3\mu}
     -g^2\Phi[2b_{\mu}^3a^{3\mu} + b_{\mu}^3b^{3\mu}],  \lb{506} \\
    & \quad  V(\Phi)=\frac{1}{2}(g^2\Phi^2+m^2\Phi).  \nonumber
\end{align}
The interaction term 
\[ - \frac{g^2}{4}[(b+a)_{\mu}\times (b+a)_{\nu}]\cdot [(b+a)^{\mu}\times (b+a)^{\nu}] \]
in $\cl_{\mathrm{inv}}(b+a)$ becomes 
\begin{align}
&-\frac{g^2}{2}(a_{\mu}^+a^{-\mu})^2 -g^2(a_{\mu}^+a^{-\mu})[(b_{\nu}^3+a_{\nu}^3)(b^{3\nu}+a^{3\nu})] \nonumber \\
& + \frac{g^2}{2}(a^{+}_{\mu}a^{+\mu})(a_{\nu}^-a^{-\nu}) +g^2[a^{+\mu}(b^{3}_{\mu}+a^{3}_{\mu})][a^{-\nu}(b^{3}_{\nu}+a^{3}_{\nu})]. \lb{507}
\end{align}
The factors $g^2\Phi^2$ and $g^2\Phi$ in Eq.(\ref{506}) come from the first and the second terms in Eq.(\ref{507}).  
From the minimum of $V(\Phi)$, we have the VEV 
\begin{equation}
     \Phi=-\frac{m^2}{2g^2},  \lb{508}
\end{equation}
and Eq.(\ref{506}) becomes 
\begin{equation}
     M^2a_{\mu}^+a^{\-\mu} + \frac{m^2}{2}[2a_{\mu}^3\ct^{\mu} + \ct_{\mu}\ct^{\mu} ].  \lb{509}
\end{equation}
If we write a tree Lagrangian for $a_{\mu}^{\pm}$ as $a_{\mu}^+(\Delta^{\mu\nu}+M^2 g^{\mu\nu})a_{\nu}^-$, 
$M^2$ is determined by the equation 
\[
     -\frac{m^2}{2g^2}= i\langle x|\mathrm{tr}\left(\Delta+M^2 \right)^{-1}|x \rangle.  
\]
Namely, although the component $a_{\mu}^3$ is massless, the components $a_{\mu}^{\pm}$ have mass $M$.  
The $m^2$ part in Eq.(\ref{509}) is $\cl_{m}$ in Eq(\ref{401}).  

     We summarize the result.  After the ghost condensation $v=g\vp_0\neq 0$, we can introduce the 
magnetic potential $\ct_{\mu}$ and the VEV $\Phi=\langle a_{\mu}^+a^{-\mu}\rangle=-m^2/(2g^2)$.  The Lagrangian $\cl_{\mathrm{inv}}$ 
gives 
\begin{align}
  \tilde{\cl}_{\mathrm{inv}}= &- \frac{1}{4}(F+H)^2+\frac{m^2}{2}[2a_{\mu}^3\ct^{\mu} + \ct_{\mu}\ct^{\mu} ] 
  + M^2a_{\mu}^+a^{-\mu} - \frac{g}{2}(F_{\mu\nu}+H_{\mu\nu})(a^{\mu}\times a^{\nu})^3 \nonumber \\
 &-\frac{g^2}{4}(a_{\mu}\times a_{\nu})^3(a^{\mu}\times a^{\nu})^3
     -\frac{1}{4}(\hat{D}_{\mu}a_{\nu}-\hat{D}_{\nu}a_{\mu})^a(\hat{D}^{\mu}a^{\nu}-\hat{D}^{\nu}a^{\mu})^a,  \lb{510}
\end{align}
where $(\hat{D}_{\mu}a_{\nu})^a=(\ptl_{\mu}a_{\nu}+gA_{\mu}^3\times a_{\nu})^a$ with $a=1,2$.  

\subsection{Lagrangian with the magnetic potential $\ct_{\mu}$}

     In the previous paper \cite{hs}, to remove the string, we performed the singular gauge transformation \cite{afg}.  
Then the Lagrangian with the massive non-Abelian magnetic potential was obtained.  
However, since we want to use the dual magnetic potential in this paper, we introduce the magnetic current $k^{\beta}$ 
as in Eq.(\ref{305}).  Namely replacing $H^{\mu\nu}$ with $H^{\mu\nu}+h_3^{\mu\nu}$, Eq.(\ref{510}) gives 
\begin{align*}
  \tilde{\cl}'_{\mathrm{inv}}= &\cl'_{\mathrm{Abel}} + \cl_{m} 
  + M^2a_{\mu}^+a^{-\mu} - \frac{g}{2}(a_{\mu}\times a_{\nu})^3(F^{\mu\nu}+H^{\mu\nu}+h_3^{\mu\nu})  \\
 &-\frac{g^2}{4}(a_{\mu}\times a_{\nu})^3(a^{\mu}\times a^{\nu})^3
     -\frac{1}{4}(\hat{D}_{\mu}a_{\nu}-\hat{D}_{\nu}a_{\mu})^a(\hat{D}^{\mu}a^{\nu}-\hat{D}^{\nu}a^{\mu})^a,  \\
   &\cl'_{\mathrm{Abel}} + \cl_{m} = -\frac{1}{4}(F + H + h_3)^2
   + \frac{m^2}{2}\left(2a^3_{\mu}\ct^{\mu} + \ct_{\mu}\ct^{\mu}\right). 
\end{align*}
The classical solution $\ct_{\mu}$ satisfies the equation of motion 
\begin{equation}
    \ptl_{\mu}(\ptl \wedge \ct)^{\mu\nu} + m^2 \ct^{\nu} -
\epsilon^{\nu\alpha\mu\beta}\frac{n_{\alpha}\ptl_{\mu}}{n^{\rho}\ptl_{\rho}}k_{\beta}=0. \lb{511}
\end{equation}
If we use it, the linear terms on $a^3_{\mu}$ disappear, and $\tilde{\cl}'_{\mathrm{inv}}$ becomes 
\begin{align}
  \tilde{\cl}'_{\mathrm{inv}}= &-\frac{1}{4}F^2 -\frac{1}{4}(\ptl\wedge \ct + h_3)^2 + \frac{m^2}{2}\ct_{\mu}\ct^{\mu} 
  + M^2a_{\mu}^+a^{-\mu} - \frac{g}{2}(a_{\mu}\times a_{\nu})^3(F^{\mu\nu}+H^{\mu\nu}+h_3^{\mu\nu}) \nonumber \\
 &-\frac{g^2}{4}(a_{\mu}\times a_{\nu})^3(a^{\mu}\times a^{\nu})^3
     -\frac{1}{4}(\hat{D}_{\mu}a_{\nu}-\hat{D}_{\nu}a_{\mu})^a(\hat{D}^{\mu}a^{\nu}-\hat{D}^{\nu}a^{\mu})^a.  \lb{512} 
\end{align}
This is the low energy effective Lagrangian with the magnetic potential $\ct_{\mu}$.  For a static magnetic charge, a solution 
$\ct_{\mu}$ which satisfies Eq.(\ref{511}) is presented in Appendix B.

\subsection{Lagrangian with the dual magnetic potential $\cc_{\mu}$}

     In Sect. 4, using the dual potential $\cc_{\mu}$, we have shown that the Lagrangian 
$\cl_{\mathrm{mAbel}}=\cl'_{\mathrm{Abel}} + \cl_{m}$ is written as in Eq.(\ref{405}).  Since $\cc_{\mu}$ 
satisfies Eq.(\ref{407}), the equality $h_2^{\mu\nu}+h_3^{\mu\nu}+h_4^{\mu\nu}=0$ holds.  We apply this 
equality to $\cl_{\mathrm{mAbel}}$ and $(a_{\mu}\times a_{\nu})^3(H^{\mu\nu}+h_{3}^{\mu\nu})$.  
Thus Eq.(\ref{512}) is rewritten as 
\begin{align}
\tilde{ \cl}'_{\mathrm{inv}}=&-\frac{1}{4}F^2
-\frac{1}{4}(\ptl \wedge \cc)^2 +\frac{m^2}{2} \cc_{\mu}\cc^{\mu} - \cc_{\mu}k^{\mu} +\Omega  \nonumber \\
 &-\frac{1}{4}(\hat{D}_{\mu}a_{\nu}-\hat{D}_{\nu}a_{\mu})^a (\hat{D}^{\mu}a^{\nu}-\hat{D}^{\nu}a^{\mu})^a + M^2 a_{\mu}^+a^{-\mu} \nonumber \\
 &- \frac{g}{2}(a_{\mu}\times a_{\nu})^3(F^{\mu\nu}+h_1^{\mu\nu}-h_4^{\mu\nu})
 -\frac{g^2}{4}(a_{\mu}\times a_{\nu})^3(a^{\mu}\times a^{\nu})^3, \lb{513}
\end{align}
where, using Eq.(\ref{211}), $(\hat{D}_{\mu}a_{\nu})^a$ becomes 
\[ (\hat{D}_{\mu}a_{\nu})^a = \ptl_{\mu}a_{\nu}^a + gf^{a3b}\left(a_{\mu}^3
+\epsilon_{\mu\alpha\beta\lambda}\frac{n^{\alpha}\ptl^{\beta}}{n^{\rho}\ptl_{\rho}}\cc^{\lambda}\right)a_{\nu}^b.  \]
Eq.(\ref{513}) is the low energy effective Lagrangian with the dual magnetic potential $\cc_{\mu}$.

\section{Magnetic charge confinement}

\subsection{The use of the Lagrangian (\ref{512})}

The classical part of the Lagrangian (\ref{512}) is 
\begin{align}
 -\frac{1}{4}(\ptl \wedge \ct+h_3)^2 + \frac{m^2}{2} \ct^{\mu}\ct_{\mu}=& 
  \frac{1}{2}\ct^{\mu}(D_{m}^{-1})_{\mu\nu}\ct^{\nu} - \ct^{\mu}{\mathcal{K}}_{\mu} \nonumber \\
 &-\frac{1}{2}k^{\mu}\frac{n^{\beta}n_{\beta}}{(n^{\rho}\ptl_{\rho})^2}
\left(g_{\mu\nu}-\frac{n_{\mu}n_{\nu}}{n_{\sigma}n^{\sigma}}\right)k^{\nu},  \lb{601}
\end{align}
where 
\begin{equation}
(D_{m}^{-1})_{\mu\nu}= g_{\mu\nu}(\square + m^2)-\ptl_{\mu}\ptl_{\nu}, \quad 
\mathcal{K}_{\mu}=\epsilon_{\mu\alpha\nu\beta}\frac{n^{\alpha}\ptl^{\nu}}{n^{\rho}\ptl_{\rho}}k^{\beta},  \lb{602}
\end{equation}
and the last term comes from $-h_3^2/4$.  
The equation (\ref{601}) can be written as 
\begin{equation}
   \frac{1}{2}\left(\ct^{\mu}-\mathcal{K}_{\sigma}D_{m}^{\sigma\mu}\right)(D_{m}^{-1})_{\mu\nu}
   \left(\ct^{\nu} - D_m^{\nu\beta}{\mathcal{K}}_{\beta} \right)
-\frac{1}{2}\mathcal{K}_{\mu}D_m^{\mu\nu}\mathcal{K}_{\nu}
-\frac{1}{2}k^{\mu}\frac{n^{\beta}n_{\beta}}{(n^{\rho}\ptl_{\rho})^2}
\left(g_{\mu\nu}-\frac{n_{\mu}n_{\nu}}{n_{\sigma}n^{\sigma}}\right)k^{\nu},  \lb{603}
\end{equation}
where 
\begin{equation}
 D_m^{\mu\nu}=\frac{g^{\mu\nu}}{\square + m^2} + \frac{\ptl^{\mu}\ptl^{\nu}}{m^2(\square+m^2)}.  \lb{604}
\end{equation}
If we apply Eq.(\ref{511}), the first term in Eq.(\ref{603}) vanishes.  
Using Eq.(\ref{a03}), $\ptl_{\nu}\mathcal{K}^{\nu}=0$ and the current conservation $\ptl_{\mu}k^{\mu}=0$, 
we find the second term in Eq.(\ref{603}) becomes 
\begin{align}
 -\frac{1}{2}\mathcal{K}_{\mu}D_m^{\mu\nu}\mathcal{K}_{\nu}=& -\frac{1}{2}\mathcal{K}_{\mu}\frac{1}{\square + m^2}\mathcal{K}^{\mu}  \nonumber \\
   =& -\frac{1}{2}k_{\mu}\frac{1}{\square + m^2}k^{\mu} 
   +\frac{1}{2}k^{\mu}\frac{\square}{\square + m^2}\frac{n^{\beta}n_{\beta}}{(n^{\rho}\ptl_{\rho})^2}
\left(g_{\mu\nu}-\frac{n_{\mu}n_{\nu}}{n_{\sigma}n^{\sigma}}\right)k^{\nu}.     \lb{605}
\end{align} 
As $\square = \square+m^2 -m^2$, Eq.(\ref{605}) is rewritten as 
\begin{align}
 -\frac{1}{2}\mathcal{K}_{\mu}D_m^{\mu\nu}\mathcal{K}_{\nu}  
   =& -\frac{1}{2}k_{\mu}\frac{1}{\square + m^2}k^{\mu} 
   -\frac{1}{2}k^{\mu}\frac{m^2}{\square + m^2}\frac{n^{\beta}n_{\beta}}{(n^{\rho}\ptl_{\rho})^2}
\left(g_{\mu\nu}-\frac{n_{\mu}n_{\nu}}{n_{\sigma}n^{\sigma}}\right)k^{\nu} \nonumber \\
& +\frac{1}{2}k^{\mu}\frac{n^{\beta}n_{\beta}}{(n^{\rho}\ptl_{\rho})^2}
\left(g_{\mu\nu}-\frac{n_{\mu}n_{\nu}}{n_{\sigma}n^{\sigma}}\right)k^{\nu}.     \lb{606}
\end{align}
The last term cancels out the third term in Eq.(\ref{603}).  Thus we obtain the magnetic current-current correlation 
\begin{align}
 \cl_{kk}=& -\frac{1}{2}\mathcal{K}_{\mu}D_m^{\mu\nu}\mathcal{K}_{\nu}
-\frac{1}{2}k^{\mu}\frac{n^{\beta}n_{\beta}}{(n^{\rho}\ptl_{\rho})^2}
\left(g_{\mu\nu}-\frac{n_{\mu}n_{\nu}}{n_{\sigma}n^{\sigma}}\right)k^{\nu} \nonumber \\
=& -\frac{1}{2}k_{\mu}\frac{1}{\square + m^2}k^{\mu} 
   -\frac{1}{2}k^{\mu}\frac{m^2}{\square + m^2}\frac{n_{\beta}n^{\beta}}{(n^{\rho}\ptl_{\rho})^2}
\left(g_{\mu\nu}-\frac{n_{\mu}n_{\nu}}{n_{\sigma}n^{\sigma}}\right)k^{\nu}.   \lb{607}
\end{align}

     If we replace the magnetic current $k_{\mu}$ with the color electric current $j_{\mu}$, Eq.({\ref{607}) becomes the 
electric current-current correlation, which was derived in the framework of the dual Ginzburg-Landau model \cite{suz, ms, sst}.  
So, by replacing electric charges with magnetic charges, we can apply the results in these references .  
For a static magnetic monopole-antimonopole pair, the current is chosen as 
\begin{equation}
     k^{\mu}(x)=Q_mg^{\mu 0}\{\delta(\vec{x}-\vec{a})-\delta(\vec{x}-\vec{b})\}, \lb{608}
\end{equation}
where the magnetic charge is $Q_m$, and the position of the monopole (antimonopole) is $\vec{a}$ ($\vec{b}$).  
We write $\vec{r}=\vec{a}-\vec{b}$, $r=|\vec{r}|$ and $n^{\mu}=(0, \vec{n})$.  The vector $\vec{n}$ is chosen as 
$\vec{n} \parallel \vec{r}$.  
\footnote{If we choose $\vec{n}$ as $\vec{n} \nparallel \vec{r}$, we obtain $V_L(\vec{r})\propto r_nK_0(m r_T)$ \cite{suz,ms}, where 
$\vec{r}=r_n\vec{n}+\vec{r}_T$ with $\vec{n}\bot \vec{r}_T$, and $r_T=|\vec{r}_T|$.  Since the modified Bessel function 
$K_0(mr_T)$ is positive and decreases rapidly as $r_T$ increases, 
the configuration with $r_T\ll 1/m$ and $r_n\cong r$ contributes to $V_L$.  
The meaning of this configuration will be discussed in the next paper.}
Then the correlation (\ref{607}) gives the monopole-antimonopole potential \cite{suz, ms, sst}
\begin{align}
  V_m(r) =& V_{Y}(r)+V_{L}(r), \quad  V_{Y}(r)= \frac{-Q_m^2}{4\pi}\frac{e^{-mr}}{r},\nonumber \\
  & 
  V_{L}(r)=\sigma r + O(e^{-mr}), \quad \sigma=\frac{Q_m^2m^2}{8\pi}\ln \left(\frac{m^2+m_{\chi}^2}{m^2}\right), 
  \lb{609}
\end{align}
where $m_{\chi}$ is the ultraviolet cutoff for the $\vec{p}_{T}$, which is the momentum component perpendicular to $\vec{r}$.  
Thus the magnetic monopoles are confined by the linear potential $V_{L}(r)$.  

     We comment on the scale $m_{\chi}$.  In the usual dual superconductor model, $m_{\chi}$ is the scale that 
the QCD-monopole condensation vanishes \cite{sst}.  In our model, since the ghost condensation 
happens at the scale $\mu_0=\Lambda e^{-4\pi^2/(\alpha_2g^2)}$ and it yields the 
mass for $\ct_{\mu}$, $m_{\chi}$ is the scale $\mu_0$.  As we showed in Ref.~\cite{hs3}, $\mu_0$ coincides with 
the QCD scale parameter $\Lambda_{\mathrm{QCD}}$.

\subsection{The use of the Lagrangian (\ref{513})}

     Next we study the classical part of Eq.(\ref{513}), i.e.,
\begin{equation}
  -\frac{1}{4}(\ptl \wedge \cc)^2 +\frac{m^2}{2} \cc_{\mu}\cc^{\mu} - \cc_{\mu}k^{\mu} +\Omega.  \lb{610}
\end{equation}
Using $\Omega$ in Eq.(\ref{408}), Eq.(\ref{610}) is rewritten as 
\begin{equation}
  \frac{1}{2}\cc^{\mu}\left\{ g_{\mu\nu}+ m^2\frac{n_{\alpha}n^{\alpha}}{(n^{\rho}\ptl_{\rho})^2}
  \left( g_{\mu\nu}-\frac{n_{\mu}n_{\nu}}{n_{\sigma}n^{\sigma}}\right)\right\}
  \left[(D_m^{-1})^{\nu\beta}\cc_{\beta}-2k^{\nu}\right]. \lb{611} 
\end{equation}
From the equation of motion (\ref{407}), $\cc^{\mu}=(D_m)^{\mu\beta}k_{\beta}$ is derived.  By 
substituting it into Eq.(\ref{611}), we find Eq.(\ref{611}) coincides with 
Eq.(\ref{607}).

     We note the second term in Eq.(\ref{607}), which yields the linear potential $V_L$, comes from 
the $h_{3\mu\nu}h_4^{\mu\nu}/4$ term in $\Omega$.  

\section{Summary and comment}

     We studied the low energy effective SU(2) gauge theory in Minkowski space.  
In the low energy region, the ghost condensation $g\vp_0 \neq 0$ happens, and 
the SU(2) symmetry breaks down to the U(1) symmetry.  
We introduced the Abelian magnetic potential $\ct_{\mu}$ as a classical solution,  
and presented the relation between $\ct_{\mu}$ and its dual potential $\cc_{\mu}$ in Minkowski space.  
It was shown that the term $h_{2}^{\mu\nu}=-\frac{1}{2}\epsilon^{\mu\nu\alpha\beta}\Lambda_{\alpha\beta}$, 
which is the Dirac string essentially, plays an important role to derive the correct Lagrangian for $\cc_{\mu}$.  

     When $g\vp_0 \neq 0$, the quantum parts of the gauge field acquire the tachyonic masses.  
These tachyonic masses are removed by the condensate 
$\langle A_{\mu}^+A^{-\mu}\rangle=\frac{1}{2}\langle(A_{\mu}^1A^{1\mu}+A_{\mu}^2A^{2\mu})\rangle$.  At the same time, 
this condensate 
makes classical parts of the gauge field massive.  Thus the magnetic potential $\ct_{\mu}$ and its dual potential 
$\cc_{\mu}$ become massive.  The effective low energy Lagrangian with $\ct_{\mu}$  is presented 
in Eq.(\ref{512}), and that with $\cc_{\mu}$ is Eq.(\ref{513}).  

     If there are static magnetic charges $Q_m$ and $-Q_m$, the classical field $\ct_{\mu}$ 
connects them.  The static potential 
between them is $V_Y(r)+V_L(r)$, where $V_Y$ is the Yukawa potential and $V_L$ is the linear potential.  
Namely the linear confining potential appears.  
If we use the dual potential $\cc_{\mu}$, the Lagrangian is not Eq. (\ref{512}) but Eq.(\ref{513}).  
However, because of the term $\Omega$ in Eq.(\ref{408}), the same static potential is obtained.  

     Usually the Dirac string is considered to be unphysical.  We cannot detect it.  In fact, 
in the massless case, the equation of motion (\ref{311}) for $\cc_{\mu}$ leads to $h_{2}^{\mu\nu}+h_{3}^{\mu\nu}=0$, and 
$h_{2}^{\mu\nu}$ disappears from the Lagrangian (\ref{309}).  
However, $h_{2}^{\mu\nu}$ is necessary to produce the correct kinetic term for the dual gauge field.  
Namely theoretical consistency requires the string term.  

     When the field $\cc_{\mu}$ becomes massive, this situation changes a little.  The equation of motion (\ref{407}) 
for $\cc_{\mu}$ leads to $h_{2}^{\mu\nu}+h_{3}^{\mu\nu}+h_{4}^{\mu\nu}=0$, and $h_{2}^{\mu\nu}$ can be removed 
from the Lagrangian (\ref{405}).  However there is the remnant $h_3^{\mu\nu}h_{4\mu\nu}/4$ in $\Omega$.  This 
term is the origin of the linear potential.

\appendix

\section{The $\epsilon$ symbol and notation}

     In this paper, we employ the metric $g_{\mu\nu}=\mathrm{diag}(1,-1,-1,-1)$.  The antisymmetric $\epsilon$ symbol 
defined by $\epsilon^{0123}=1$ satisfies the formulae 
\begin{equation}
    \epsilon^{\alpha\beta\rho\sigma}\epsilon_{\alpha\lambda\mu\nu}=-\left|
\begin{array}{ccc}
 \delta^{\beta}_{\lambda}&\delta^{\beta}_{\mu} & \delta^{\beta}_{\nu} \\
 \delta^{\rho}_{\lambda}&\delta^{\rho}_{\mu} & \delta^{\rho}_{\nu} \\
\delta^{\sigma}_{\lambda} & \delta^{\sigma}_{\mu}& \delta^{\sigma}_{\nu}
\end{array}
\right|,\quad
     \epsilon^{\alpha\beta\rho\sigma}\epsilon_{\rho\sigma\mu\nu}=-2(\delta^{\alpha}_{\mu}\delta^{\beta}_{\nu}
     -\delta^{\alpha}_{\nu}\delta^{\beta}_{\mu}).   \lb{a01} 
\end{equation}
From Eq.(\ref{a01}), the following relations are obtained: 
\begin{align}
     \epsilon_{\mu\nu\alpha\beta}\frac{n^{\alpha}J^{\beta}}{n^{\rho}\ptl_{\rho}}
     \epsilon^{\mu\nu\kappa\lambda}\frac{n_{\kappa}K_{\lambda}}{n^{\sigma}\ptl_{\sigma}}
     =&
2J^{\mu}\frac{n_{\beta}n^{\beta}}{(n^{\rho}\ptl_{\rho})^2}\left(g_{\mu\nu}-\frac{n_{\mu}n_{\nu}}{n_{\sigma}n^{\sigma}}\right)K^{\nu},  \lb{a02} \\
    \epsilon_{\mu\nu\alpha\beta}\frac{n^{\nu}\ptl^{\alpha}J^{\beta}}{n^{\rho}\ptl_{\rho}}
\epsilon^{\mu\eta\kappa\lambda}\frac{n_{\eta}\ptl_{\kappa}K_{\lambda}}{n^{\sigma}\ptl_{\sigma}}
 =& J^{\mu}\frac{1}{(n^{\rho}\ptl_{\rho})^2} \big\{ (n^{\sigma}\ptl_{\sigma})^2 g_{\mu\nu}-n^{\sigma}n_{\sigma}\Box g_{\mu\nu}
 + n_{\sigma}n^{\sigma}\ptl_{\mu}\ptl_{\nu}  \nonumber \\
 &  + n_{\mu}n_{\nu}\Box - n^{\sigma}\ptl_{\sigma}\ptl_{\mu}n_{\nu}-n^{\sigma}\ptl_{\sigma}n_{\mu}\ptl_{\nu} \big\} K^{\nu}.  \lb{a03} 
\end{align}

     For simplicity, we use the notations 
\begin{equation}
 (\ptl\wedge \cc)^{\mu\nu}=\ptl^{\mu}\cc^{\nu}-\ptl^{\nu}\cc^{\mu},\quad 
     \ptl_{\sigma}(\ptl\wedge \cc)^{\sigma\nu}=\Box \cc^{\nu}-\ptl^{\nu}\ptl_{\sigma}\cc^{\sigma}  \lb{a04}
\end{equation}
and 
\[ H^2 = H_{\mu\nu}H^{\mu\nu},  \]
where $H_{\mu\nu}$ is an antisymmetric tensor.

\section{Monopole solutions and dual solutions}

     In this appendix, we present monopole solutions and dual solutions for a static magnetic charge 
\begin{equation}
     k^{\beta}=\delta^{\beta}_0 \frac{4\pi N}{g} \delta(x)\delta(y)\delta(z), \quad (N=\pm 1,\pm 2, \cdots).  \lb{b01}
\end{equation}
The term $\Lambda^{\mu\nu}$ defined in Eq.(\ref{207}) is discussed.

\subsection{The massless case}

      Choosing the Dirac string on the negative $z$-axis, it is shown that the monopole solution 
\begin{equation}
     \ct_{\mu}=\frac{N}{g}(\cos \theta -1)\ptl_{\mu}\phi=\frac{N}{g}\frac{z-r}{r\rho^2}(0,-y,x,0)   \lb{b02}
\end{equation}
satisfies the equation 
\begin{equation}
    \ptl_{\mu}(\ptl \wedge \ct)^{\mu\nu} -
\epsilon^{\nu\alpha\mu\beta}\frac{n_{\alpha}\ptl_{\mu}}{n^{\rho}\ptl_{\rho}}k_{\beta}=0,  \lb{b03}
\end{equation}
where $(r, \theta, \phi)$ are spherical coordinates, $\rho=\sqrt{x^2+y^2}$, 
and the space-like vector is chosen as $n^{\rho}=\delta^{\rho}_3$.  
The corresponding dual potential satisfies the equation 
\begin{equation}
     \ptl_{\mu}(\ptl \wedge \cc)^{\mu\nu}-k^{\nu}=0,  \lb{b04}
\end{equation}
and the solution is  
\begin{equation}
     \cc^{\mu}=\frac{N}{g}\frac{1}{r}\delta^{\mu}_{0}.  \lb{b05}
\end{equation}

     Now, using $\ct^{\mu}$, we calculate $\hh^{\mu\nu}$.  Substituting Eq.(\ref{b02}) into Eq.(\ref{202}), we find 
\begin{equation}
     \hh^{0j}=-\frac{N}{g}\frac{x^j}{r^3},\ (j=1,2),\quad 
     \hh^{03}=-\frac{N}{g}\frac{x^3}{r^3}-\frac{4\pi N}{g}\theta(-z)\delta(x)\delta(y),  \lb{b06}
\end{equation}
where $(\ptl_{x}^2+\ptl_{y}^2)\ln \rho=2\pi \delta(x)\delta(y)$ has been used.  As 
$\nabla^2(1/r)=-4\pi \delta(\mathbf{r})$, it is easy to show $\ptl_{\nu}\hh^{\mu\nu}=\ptl_{j}\hh^{0j}=0$.  
Namely Eq.(\ref{205}) is satisfied.  

    Next, to calculate $\hh^{\mu\nu}$, we use the dual potential $\cc^{\mu}$.  
From Eqs.(\ref{208}) and (\ref{b05}), $\hh^{\mu\nu}$ becomes  
\begin{equation}
      \hh^{0j}=-\ptl^j\cc^0=-\frac{N}{g}\frac{x^j}{r^3},\ (j=1,2),\quad 
       \hh^{03}=-\frac{N}{g}\frac{x^3}{r^3}+\Lambda^{03}.  \lb{b07}
\end{equation}
We follow the Zwanziger's  definition \cite{zwa} 
\begin{equation}
     \frac{1}{\ptl_z}f(x,y,z)=a\int_0^{\infty} f(x,y,z-s)ds -(1-a)\int_0^{\infty}f(x,y,z+s)ds,  \lb{b08}
\end{equation}
and, to put the Dirac string on the nagative $z$-axis, we set $a=0$.   
This choice gives 
\begin{equation*}
      \frac{1}{\ptl_z}\delta(z)=-\theta(-z),  \lb{b08}
\end{equation*}
and we find 
\[  \Lambda^{03}=\frac{1}{\ptl_z}(\Box \cc^0)= -\frac{4\pi N}{g}\theta(-z)\delta(x)\delta(y).  \]
Thus $\Lambda^{\mu\nu}$ term represents the Dirac string part, and Eq.(\ref{b07}) coincides with Eq.(\ref{b06}).

\subsection{The massive case}

     In this case, $\ct^{\mu}$ fulfills the equation 
\begin{equation}
    \ptl_{\mu}(\ptl \wedge \ct)^{\mu\nu} + m^2\ct^{\nu} -
\epsilon^{\nu\alpha\mu\beta}\frac{n_{\alpha}\ptl_{\mu}}{n^{\rho}\ptl_{\rho}}k_{\beta}=0. \lb{b09}
\end{equation}
By modifying Eq.(\ref{b02}), we find Eq.(\ref{b09}) is satisfied by the solution 
\begin{equation}
     \ct_{\mu}=\frac{N}{g}\frac{z-r}{r\rho^2}e^{-mr}(0,-y,x,0).   \lb{b10}
\end{equation}

     In the same way, the dual potential 
\begin{equation}
     \cc^{\mu}=\frac{N}{g}\frac{e^{-mr}}{r}\delta^{\mu}_{0}  \lb{b11}
\end{equation}
fulfills the equation 
\begin{equation}
     \ptl_{\mu}(\ptl \wedge \cc)^{\mu\nu} + m^2\cc^{\nu} -k^{\nu}=0.   \lb{b12}
\end{equation}

\section{Ghost condensation in Minkowski space}

     At the one-loop level, the Lagrangian (\ref{504}) yields the effective potential for $v=g\vp_0$  as  
\begin{equation}
     V_M(v)=\frac{v^2}{2\alpha_2 g^2} + iV_{\mathrm{gh}}(v),\quad V_{\mathrm{gh}}(v)=
     \int \frac{d^4p}{(2\pi)^{4}} \ln [(-p^2)^2+v^2].  \lb{c01}
\end{equation}
In Ref.~\cite{hs2}, we calculated $V_{\mathrm{gh}}$ directly, and showed Eq.(\ref{c01}) becomes 
\begin{equation}
     V_M(v)=\frac{v^2}{2\alpha_2 g^2} -i\frac{v^2}{32\pi}.  \lb{c02}
\end{equation}
Therefore the condition $V_M'(v)=0$ gives $v=0$.  

     However, when $v=0$, the integrand $\ln [(-p^2)^2+v^2]$ diverges at $p^2=0$, and the calculation in 
Ref.~\cite{hs2} is inapplicable.  
So we should replace $p^2$ with $p^2+i\epsilon$ as usual, and set $\epsilon \to 0$ after the $p$-integration.  
Thus we consider 
\begin{equation}
     iV_{\mathrm{gh}}(v)=
     i \int \frac{d^4p}{(2\pi)^{4}} \ln [(-p^2-i \epsilon -iv)(-p^2-i\epsilon  +iv)].  \lb{c03}
\end{equation}
When $\epsilon < v$, if we take the limit $\epsilon \to 0$, the result (\ref{c02}) is obtained.  In the 
$p^0$ plane, since the pole $\{|\vec{p}|^2-i \epsilon + iv\}^{1/2}$ ($-\{|\vec{p}|^2-i \epsilon + iv\}^{1/2}$) is 
in the first quadrant (the third quadrant), the usual Wick rotation is inapplicable.  On the other hand, 
when $\epsilon > v$, the poles $\{|\vec{p}|^2-i \epsilon \pm iv\}^{1/2}$ are in the fourth quadrant, and 
$ -\{|\vec{p}|^2-i \epsilon \pm iv\}^{1/2}$ 
are in the second quadrant.  Then we can apply the usual Wick rotation, which is performed by 
the replacement $p^0 \to -ip_4$ and $\int dp^0 \to i \int dp_4$.  After that, we can take the limit $\epsilon \to 0$, and 
we find 
\begin{equation}
     V_M(v)=\frac{v^2}{2\alpha_2 g^2} -\int \frac{d^4p_E}{(2\pi)^{4}} \ln [(p_E^2 -iv)(p_E^2 +iv)], \lb{c04}
\end{equation}
where $(p_E)_{\mu}=(\vec{p},p_4)$ is the Euclidean four-momentum.  This is the usual Euclidean 
potential, and its minimum is at $v\neq 0$ \cite{hs3}.  Namely the ghost condensation happens in 
Minkowski space as well.  

\begin{figure}
\begin{center}
\includegraphics{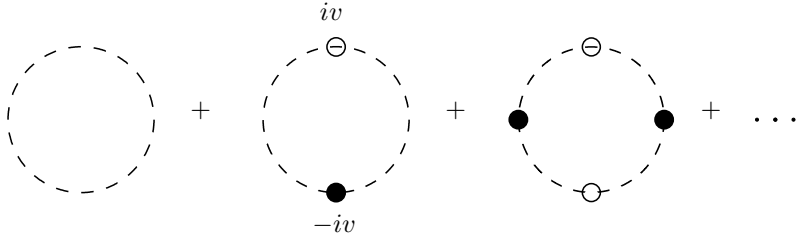}
\caption{The one-loop ghost diagrams.  The dashed line is the ghost propagator $\langle c^{\mp}\cb^{\pm}\rangle$, and 
the blobs represent $\pm iv$.}
\label{fig1}
\end{center}
\end{figure}

     We make a comment.  The one-loop diagrams in Fig.C1 lead to the series 
\begin{equation}
    \int \frac{d^4p}{(2\pi)^{4}i} \ln(-p^2 -i\epsilon)^2 + 
     \sum_{n=1}^{\infty} \frac{-1}{n}\int \frac{d^4p}{(2\pi)^{4}i} \left\{-\frac{v^2}{(-p^2-i\epsilon)^2}\right\}^n.  \lb{c05}
\end{equation}
Under the condition 
\begin{equation}
     \left| \frac{v^2}{(p^2+i \epsilon)^2} \right|<1,  \lb{c06}
\end{equation}
this series converges as 
\begin{equation}
    -i \int \frac{d^4p}{(2\pi)^{4}} \ln(-p^2 -i\epsilon)^2  
     -i \int \frac{d^4p}{(2\pi)^{4}} \ln \left[1+\frac{v^2}{(-p^2-i\epsilon)^2}\right].  \lb{c07}
\end{equation}
This expression gives the potential $V_{\mathrm{gh}}$.  
For an arbitrary value of $p^2$, the condition (\ref{c06}) is satisfied if $\epsilon > v$.  Namely, 
the condition $\epsilon > v$ is required for convergence.

\end{document}